\documentclass[aip,twocolumn,amsmath,amssymb,letterpaper]{revtex4}

\pdfoutput=1
\usepackage{hyperref}
\usepackage{graphicx}

\def\cal#1{\mathcal{#1}}
\def\eqq#1{Eq.~(\ref{#1})}

\def\av#1{\langle #1 \rangle}

\def\beq{\begin{equation}}
\def\eeq{\end{equation}}
\def\bea{\begin{eqnarray}}
\def\eea{\end{eqnarray}}

\def\kt{k_{\rm B}T}

\begin{document}

\title{Limit of validity of Ostwald's rule of stages in a statistical mechanical model of crystallization}

\author{Lester O. Hedges and Stephen Whitelam\footnote{\tt{swhitelam@lbl.gov}}} 
\affiliation{Molecular Foundry, Lawrence Berkeley National Laboratory, 1 Cyclotron Road, Berkeley, CA 94720, USA}

\begin{abstract}
We have only rules of thumb with which to predict how a material will crystallize, chief among which is Ostwald's rule of stages. It states that the first phase to appear upon transformation of a parent phase is the one closest to it in free energy. Although sometimes upheld, the rule is without theoretical foundation and is not universally obeyed, highlighting the need for microscopic understanding of crystallization controls. Here we study in detail the crystallization pathways of a prototypical model of patchy particles. The range of crystallization pathways it exhibits is richer than can be predicted by Ostwald's rule, but a combination of simulation and analytic theory reveals clearly how these pathways are selected by microscopic parameters. Our results suggest strategies for controlling self-assembly pathways in simulation and experiment.

\end{abstract}
\maketitle

\section{Introduction}

Crystallization frequently happens in a  `multi-stage' manner, with a parent phase (e.g. a solution) first transforming into an intermediate phase (e.g. a dense liquid) before the stable solid emerges~\cite{zhang2009nucleation,chung2008multiphase,chung2010self}. One of the few guidelines we have for predicting when crystallization intermediates will appear is Ostwald's rule of stages, which states that the parent phase will first transform into the phase closest to it in free energy~\cite{ostwald1897studies,threlfall2003structural}. It is widely upheld. For instance, sulfur crystallizes from solution by first forming a dense liquid~\cite{ostwald1897studies}. Melts~\cite{cech1957evidence} and aerosols~\cite{fox1995metastable} also display expected precursors of the stable crystal. On the computer, a microscopic analog of the rule is seen: the freezing of polar fluids, model proteins~\cite{wolde1999homogeneous} and molecular nitrogen~\cite{leyssale2003molecular} can all take place via nuclei whose composition reflects that of an intermediate phase. But the rule has no theoretical foundation~\cite{cardew1984kinetics} and is not universally obeyed. Amino acid crystallization~\cite{kitamura2009strategy}, the simulated freezing of molecular CO$_2$~\cite{leyssale2005molecular} and Potts model phase transformations~\cite{sanders2007competitive} can all take place {\em without} involvement of metastable polymorphs. Further, simulations of charged colloids~\cite{sanz2007evidence} show that sluggish dynamics can invalidate the closely-related Stranski-Totomanow conjecture~\cite{stranski1933rate}, the prediction that the first phase seen is the one separated from the parent phase by the smallest free energy barrier.

The limitations of these rules of thumb motivate us to look for connections between crystallization pathways and {\em microscopic} features of particle interactions and dynamics. Here we study a lattice model of anisotropic particles. Our model is designed to mimic, in a generic way, the ability of materials such as proteins~\cite{galkin2000control,chung2010self} and ions~\cite{radha2010transformation} to crystallize by first forming a {\em disordered} phase. It is also designed to be simple enough to allow thorough assessment of how its microscopic parameters control its crystallization behavior. Here we describe this behavior in detail. We show where in phase space Ostwald's rule is likely to hold, and where it is likely to fail. Because the essence of the microscopic features of our model are common to a wide range of physical systems, the trends we have identifed might be used as a guide to select particular crystallization pathways in simulation and experiment.

\section{Model and simulation protocol} We consider a collection of particles that live on a featureless two-dimensional substrate, which we model as a square lattice. Lattice sites may be vacant or be occupied by a particle. Nearest-neighbor particles receive a `nonspecific' interaction energy reward of $-J$. Particles are anisotropic, and can point in any of $R$ discrete directions. Neighboring particles receive an additional `specific' energy reward of $-Q$ if they are aligned, and a penalty of $+Q$ if they are antialigned. The larger is $R$, the more precisely must two particles align before they receive the specific binding reward (we shall explore the effect of varying $R$). Each particle on the substrate feels a chemical potential $-\mu$. 

We simulated our model using the Monte Carlo procedure described in Appendix A. This procedure allows particles to translate (adsorb to and desorb from the substrate), and to rotate in place on the substrate. We have explored the effect of varying extensively the relative rate $r$ of proposing rotation and translation moves, because we expect rotational and translational mobilities to vary considerably from one material to another. For instance, the limit of slow rotations ($r \ll 1$) might be appropriate for particles, like DNA-linked colloids, that must unbind in order to rotate appreciably~\cite{dai2010universal}. It is also likely that particles of different sizes explore their positional and angular interaction ranges at different rates~\cite{note}. We used this procedure in concert with umbrella sampling~\cite{torrie1977nonphysical} to calculate free energy landscapes for crystallization (biasing the size and degree of crystallinity of the system's largest cluster~\cite{wolde1999homogeneous}), and in concert with transition path sampling~\cite{bolhuis2002transition,pan2004dynamics} and forward flux sampling~\cite{allen2005sampling} to generate dynamical crystallization trajectories. The combination of these methods reveals the distinct effect on crystallization of thermodynamics and dynamics.

\section{Simulation Results} 
\begin{figure}[!t]
\includegraphics[width=\linewidth]{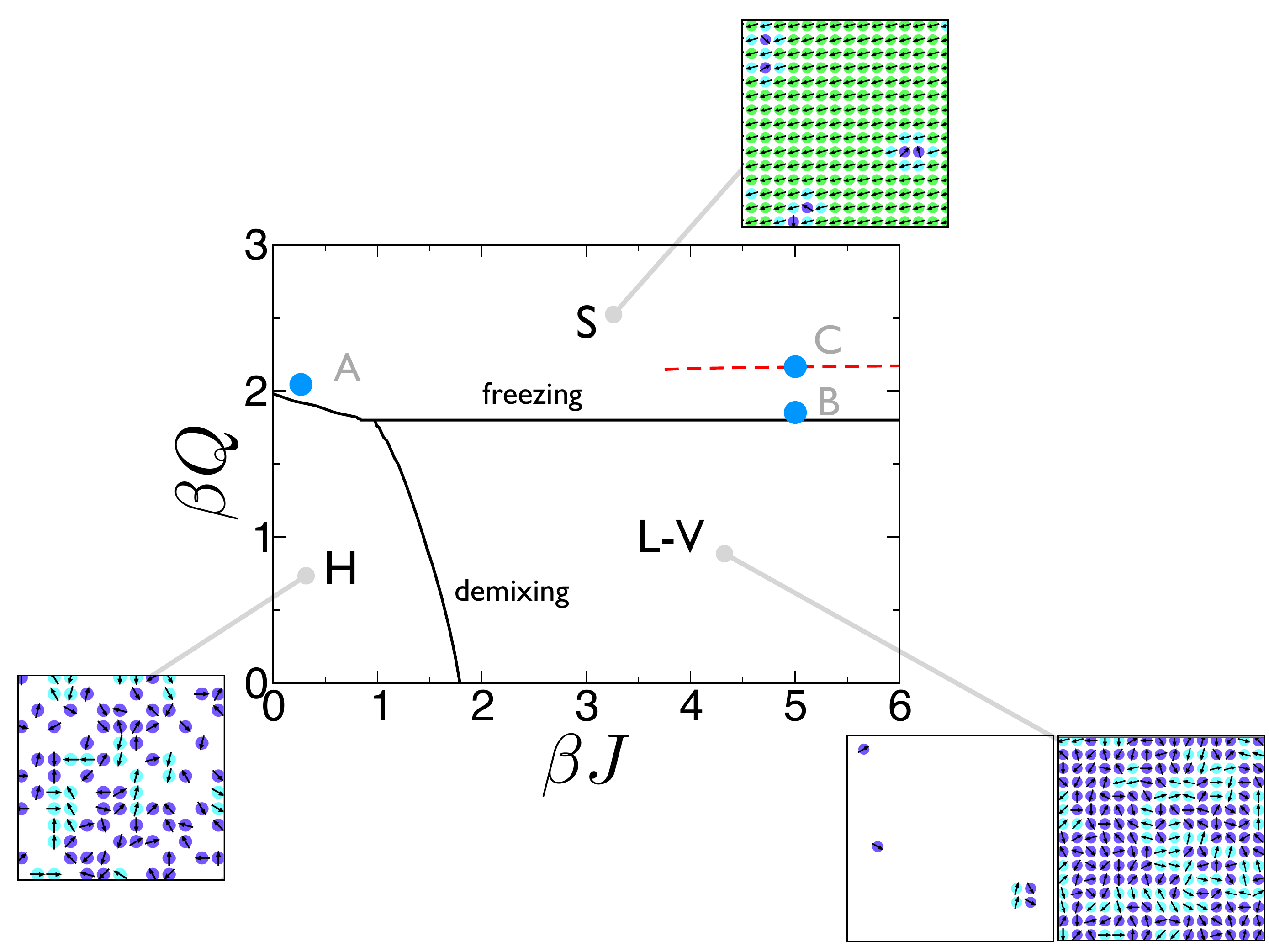}
\caption{\label{fig1} Model phase diagram in the space of nonspecific $(J)$ and specific $(Q)$ attraction strengths, with phase snapshots, for fixed chemical potential (see Appendix). When both interactions are weak the stable phase is a homogeneous fluid H of moderate density and no orientational order. `Demixing' into a sparse vapor V and a dense liquid L (both orientationally disordered) can be induced by application of a strong nonspecific attraction, $J$ (the critical point on the horizontal axis is the regular Ising model one). The orientationally-ordered solid phase S is stable when the specific interaction $Q$ is large enough, above the freezing line. We are interested in the region of solid stability to the right of the demixing line. There, the sparse vapor phase lies above the dense fluid phase in free energy, which in turn lies above the stable solid phase. Ostwald's rule suggests that if we start from the vapor phase then the liquid will emerge prior to crystallization. But as we describe below, this is not always true. The thermodynamically-favored critical nucleus is liquidlike below the horizontal red line, but crystalline above it. Points A, B, and C correspond to the simulations shown in Fig.~\ref{fig2}. The connection between this phase diagram and the conventional temperature-density one is given at mean-field level in Fig.~\ref{figa6}.}
\end{figure}

In Fig.~\ref{fig1} we show the phase diagram of the model as a function of interaction strengths $J$ and $Q$, for angular specificity $R=24$. Snapshots of phases are also shown: particles with 4 parallel neighbors (crystalline particles) are green; particles with no parallel neighbors (fluid particles) are dark blue. Intermediate particles, with 1--3 parallel neighbors, are blue-green. When both interactions are weak the stable phase is a homogeneous fluid phase (H) of moderate density and little orientational order. When $Q$ is large enough (above the `freezing line'), the orientationally-ordered solid (S) is stable. When $J$ is sufficiently large (to the right of the `demixing line'), phase H disappears and dense liquid (L) and sparse vapor (V) phases become viable. Our choice of $\mu$ (see Appendix) ensures that above the freezing line and to the right of the demixing line the liquid phase lies intermediate in free energy between the vapor phase and the stable solid. In this regime, Ostwald's rule suggests that a vapor should transform into a liquid before it crystallizes. But is this true?

\begin{figure*}[!t]
\includegraphics[width=0.93\linewidth]{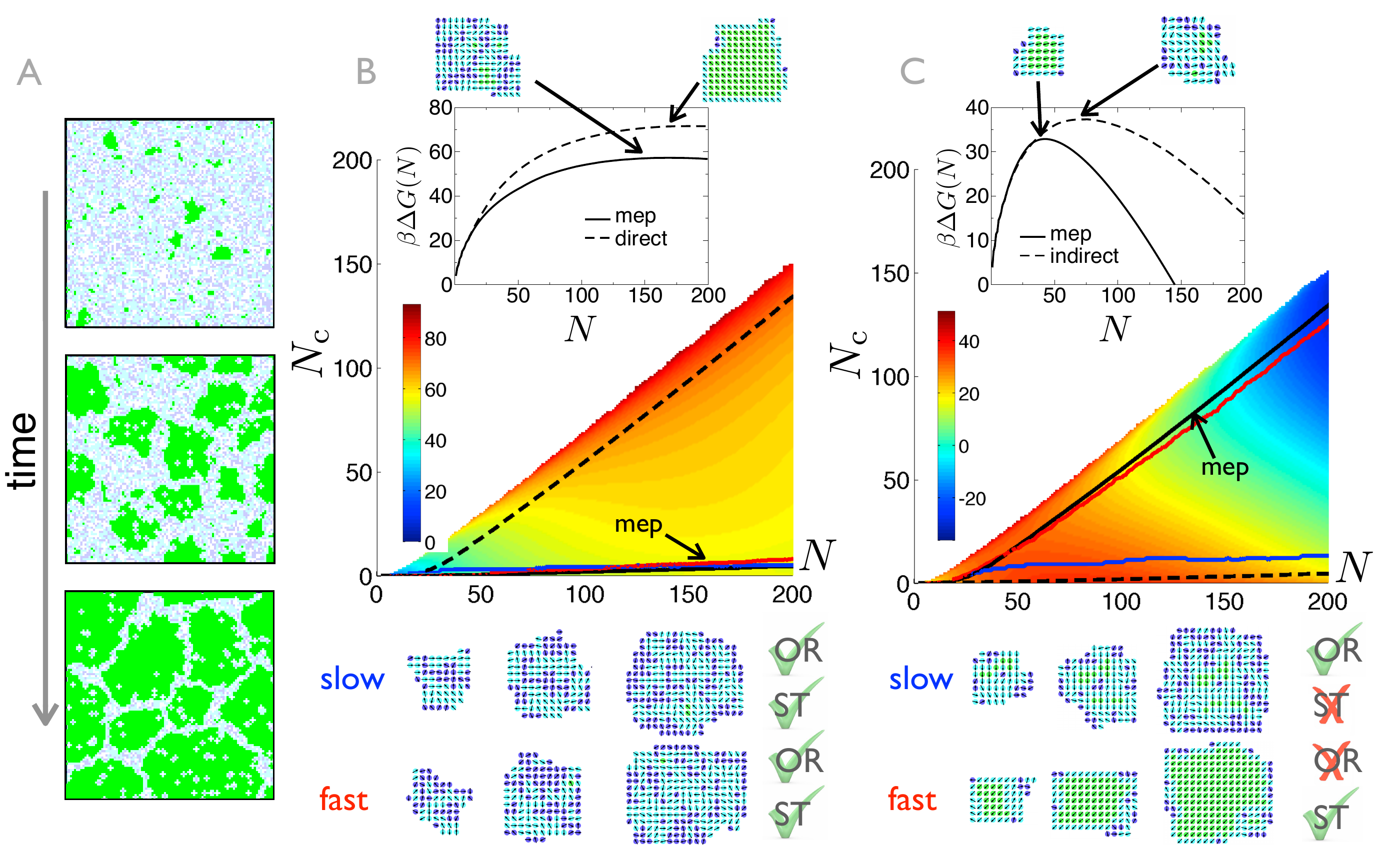}
\caption{\label{fig2} Crystallization pathways at points A, B and C in phase space of Fig.~\ref{fig1}. At point A, only phases H and S are viable, and crystallization is straightforward: it consists of the direct transformation of H into S. Points B and C lie in the regime of interest, where the vapor, liquid and crystal phases lie in descending order of free energy. We show free energy surfaces (in a space of cluster size $N$ and degree of crystallinity $N_{\rm c}$) and dynamical trajectories for crystallization at phase points B and C, between which the thermodynamic mechanism for crystallization switches from an `indirect' one to a `direct' one. Snapshots bottom right show configurations generated using different dynamical protocols. The resulting trajectories sometimes uphold and sometimes violate Ostwald's rule (OR) and the Stranski-Totomanow  (ST) conjecture.}
\end{figure*}

Crystallization pathways at points A, B and C in phase space are shown in Fig.~\ref{fig2}. At point A in phase space, only phases H and S are viable, and crystallization is straightforward: it consists of the direct transformation of H into S. At point B on the phase diagram (in our regime of interest) an empty substrate immediately becomes host to the metastable low-density vapor. We show in Fig.~\ref{fig2} the free energy of formation from the vapor of a dense nucleus as a function of nucleus size $N$ and the number of crystalline particles $N_{\rm c}$ it contains. The minimum energy pathway (`mep', solid line) from the vapor to the crystal is an {\em indirect} one (sometimes called a `two-step' or `nonclassical' pathway) that displays a liquidlike critical nucleus. The direct pathway (dashed line) via a crystalline critical nucleus is disfavored by about 15 $\kt$. The indirect pathway is made possible by the intermediate liquid phase, but because we are far from the demixing line is not a result of critical density fluctuations~\cite{wolde1997epc}. Dynamical trajectories generated using a wide range of particle rotation rates adhere to the indirect pathway (red line, fast rotation: $r=99$; blue line, slow rotation: $r=0.01$): a liquid nucleates on the substrate, and only subsequently does the crystal emerge from the liquid. For rapid rates of rotation the postcritical liquid readily transforms into a crystal while still only of small size \href{http://nanotheory.lbl.gov/people/mwnn_paper/point_B_fast.mp4}{M1}, while for sluggish rotation rates the liquid consumes the substrate, and fails to crystallize during the course of the simulation \href{http://nanotheory.lbl.gov/people/mwnn_paper/point_B_slow.mp4}{M2}. 

At point C in phase space the liquid remains intermediate in free energy between the parent phase and the stable solid, but the driving force for crystallization is qualitatively different: the {\em direct} pathway, with a crystalline critical nucleus, is preferred! The indirect pathway with a liquidlike critical nucleus is still viable, but is disfavored by about $5\, \kt$. Because of this relatively small discrepancy in barrier heights, both pathways can be seen in dynamic simulations. For a sufficiently fast rotation rate (red line, $r=99$) the direct pathway is taken: a crystal nucleates and grows on the substrate. No liquid is seen. For a slow rotation rate (blue line, $r=0.01$) the indirect pathway {\em is} seen, and the substrate is again consumed by a liquid, \href{http://nanotheory.lbl.gov/people/mwnn_paper/point_C_fast.mp4}{M3}, \href{http://nanotheory.lbl.gov/people/mwnn_paper/point_C_slow.mp4}{M4}, and~\ref{figa3}. 

A relic of the liquid phase therefore influences the crystallization pathway some way past the freezing line, but eventually thermodynamics favors a direct mode of crystallization: the critical nucleus is crystalline above the horizontal red line on the phase diagram of Fig.~\ref{fig1}. The question of how much liquid is seen in simulations is one that cannot be addressed by Ostwald's rule (`OR' in caption). Assuming that it applies (it pertains to metastable intermediate phases, and the liquid at B and C is at most weakly so), it is satisfied in a macroscopic sense by the `slow rotation' (blue) trajectories at point B, but only in a microscopic sense by the `fast rotation' (red) trajectories. It is violated by the `fast' trajectories at C, but not by the `slow' trajectories. Moreover, the latter involve passage over a free energy barrier larger than the smallest available, going against the sense of the Stranski-Totomanow conjecture (`ST' in caption). 

\begin{figure*}[!htb]
\includegraphics[width=\linewidth]{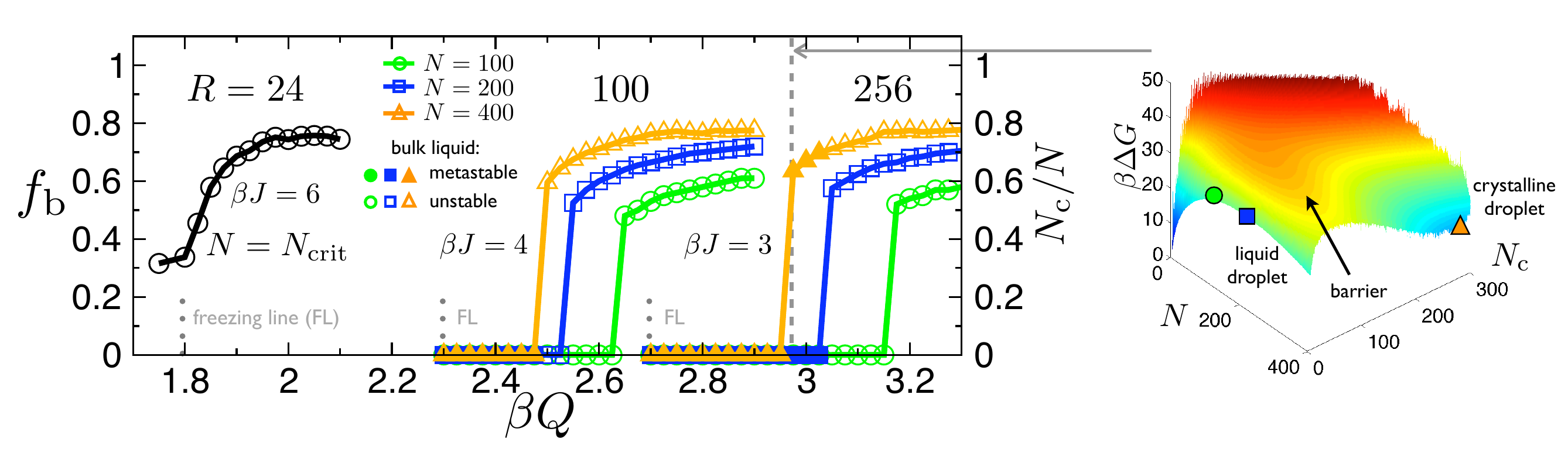}
\caption{\label{fig3} Microscopic thermodynamic controls of crystallization. Thermodynamically-preferred composition ($f_{\rm b}$, left, is fraction of crystalline interactions, and $N_{\rm c}$, right, is number of crystalline particles) of a critical droplet $N_{\rm crit}$ or one of specified size $N$, as a function of specific interaction strength $Q$, for three values of angular specificity $R$. The {\em bulk} solid is stable to the right of each freezing line (vertical gray dots labeled `FL'), but a droplet of {\em finite} size prefers to be liquid some way past the phase boundary. As $Q$ increases, however, small droplets prefer to be solid, a trend that opposes Ostwald's rule. By contrast, increasing $R$ acts in favor of Ostwald's rule by rendering the liquid more strongly metastable to crystallization: corresponding free energy surfaces (at right) possess barriers that hinder the crystallization of a liquid droplet.}
\end{figure*}

The microscopic thermodynamic control of our model's crystallization pathway is the competition between the angular specificity $R$ and potency $Q$ of specific binding. In Fig.~\ref{fig3} we show that the thermodynamically-preferred composition of the critical nucleus (and of small nuclei of specified size) is liquidlike for small $Q$ (just above the freezing line), and becomes increasingly crystalline as one increases $Q$: this trend acts {\em against} the sense of Ostwald's rule. By contrast, increasing $R$ acts in {\em favor} of Ostwald's rule: as $R$ grows, the liquid becomes more strongly metastable to crystallization, giving rise to free energy barriers that hinder the transformation of a liquid droplet into a solid one (compare surfaces top right and top left). 

\section{Analytic Results} Interestingly, we can anticipate these thermodynamic trends using simple microscopic theory. We calculated analytically (see Appendix B) the model's bulk free energy $f_{\rm eff}$, in a self-consistent mean-field approximation, as a function of order parameters $\rho$ (density) and $\tau$ (crystallinity):
\begin{widetext}
\begin{equation}
\label{eq1}
f_{\rm eff}(\rho,\tau)=\frac{1}{2} \left(J z\rho^2+Qz \tau^2\right) 
-\kt \ln \left( 1+ {\rm e}^{\beta (Jz  \rho+\mu)} \left[ R-2+2 \cosh(\beta Qz \tau) \right]\right).
\end{equation}
\end{widetext}

\begin{figure*}[!t]
\includegraphics[width=\linewidth]{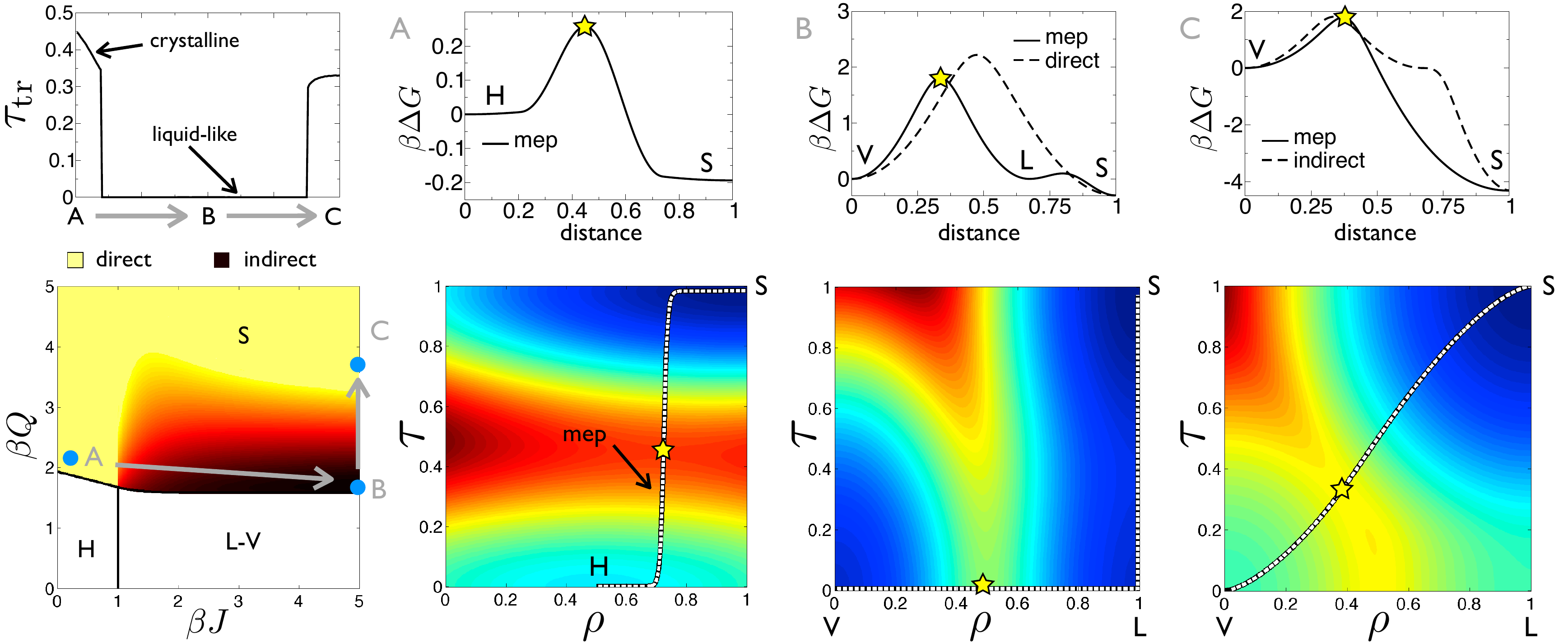} 
\caption{\label{fig4} Visualizing the shaping of crystallization landscapes by microscopic parameters. Mean-field `pathway diagram' (bottom left, $R=24$) and free energy surfaces at points A, B, and C in phase space (mep is dotted white line; star denotes transition state), calculated using~\eqq{eq1}. Above each surface we show free energy along the mep (solid black line with star), and, for comparison, free energy along a direct pathway (B) and an indirect pathway (C) (shown dotted; these pathways are not shown in panels below). Pathway diagram shading indicates the preferred crystallization mechanism; the plot above it shows where on the lines A $\to$ B $\to$ C the transition state $\tau_{\rm tr}$ is crystalline. The qualitative trends identified mirror those seen in our simulations.}
\end{figure*}

Here $z=4$ is the lattice coordination number. From \eqq{eq1} we obtained the phase diagram shown in Fig.~\ref{fig4}, which resembles qualitatively its simulated counterpart. The microscopic parameters of the model control distinct critical behaviors, but they also shape the bulk free energy landscape even in regimes {\em away} from any phase transition.  Furthermore, previous work reveals that the bulk free energy landscape suggests the qualitative character of the thermodynamically-preferred crystallization pathway~\cite{shen1996bcc, lutsko2006ted}: bulk wells indicate driving forces for appearance of phases, and bulk barriers are an important component of droplet surface tensions in a Cahn-Hilliard approximation. On bulk surfaces we calculated the minimum energy pathways between parent and solid phases (shown by white dashed lines on surfaces at points A, B, C), and shaded the phase diagram according to how `direct' are those pathways. The resulting `pathway diagram' describes qualitatively the thermodynamic driving force for crystallization for given microscopic parameters, and mirrors the changes of cluster composition seen in our umbrella sampling (thermodynamic) simulations: indirect crystallization pathways become viable to the right of the demixing line, and are supplanted by a direct mechanism some distance above the freezing line. This correspondence is a central result of this paper. This change of mechanism takes place in a regime of phase space in which the hierarchy of stable phase remains unchanged: in other words, the {\em phase} diagram in this regime is featureless, but the {\em pathway} diagram is not.

\begin{figure}[!b]
\includegraphics[width=\columnwidth]{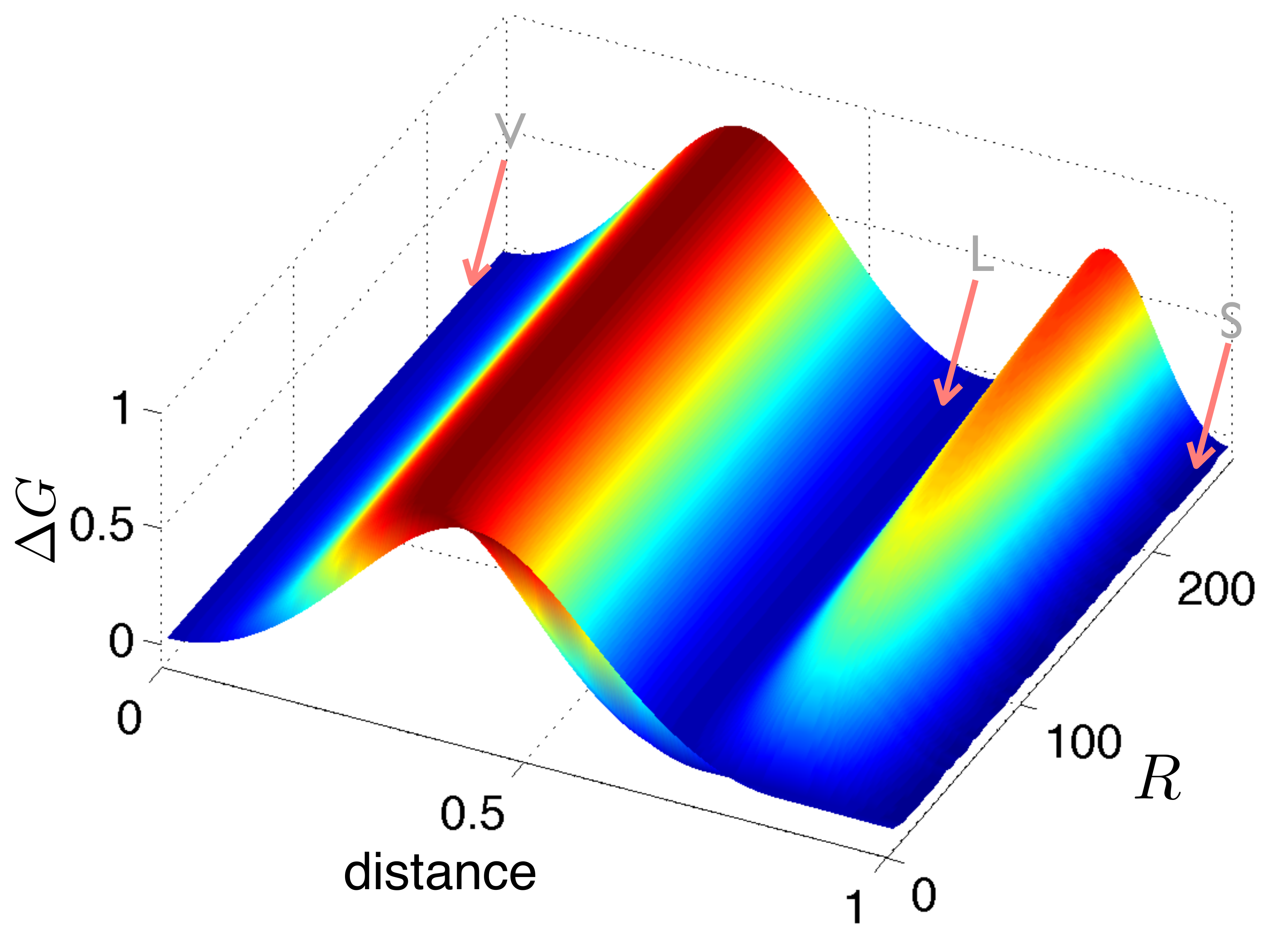} 
\caption{\label{fig5} Effect of varying $R$, the number of accessible rotational states. We plot free energy (vertical) along the minimum energy pathway (`distance' axis) of a mean-field free energy surface at $J=3\, \kt$, for different values of $R$. The value of $Q$ is such that we sit just above the freezing line for each $R$ (see Fig.~\ref{fig4}, for $R=24$). In all cases the pathway observed is an indirect one from vapor (V) to liquid (L) to solid (S). The first barrier seen corresponds to the vapor-to-liquid transformation, and the second to the liquid-to-crystal transformation. Increasing $R$ leads to the growth of a large entropic barrier to crystallization.} 
\end{figure}

\begin{figure*}[htb]
\includegraphics[width=0.95\linewidth]{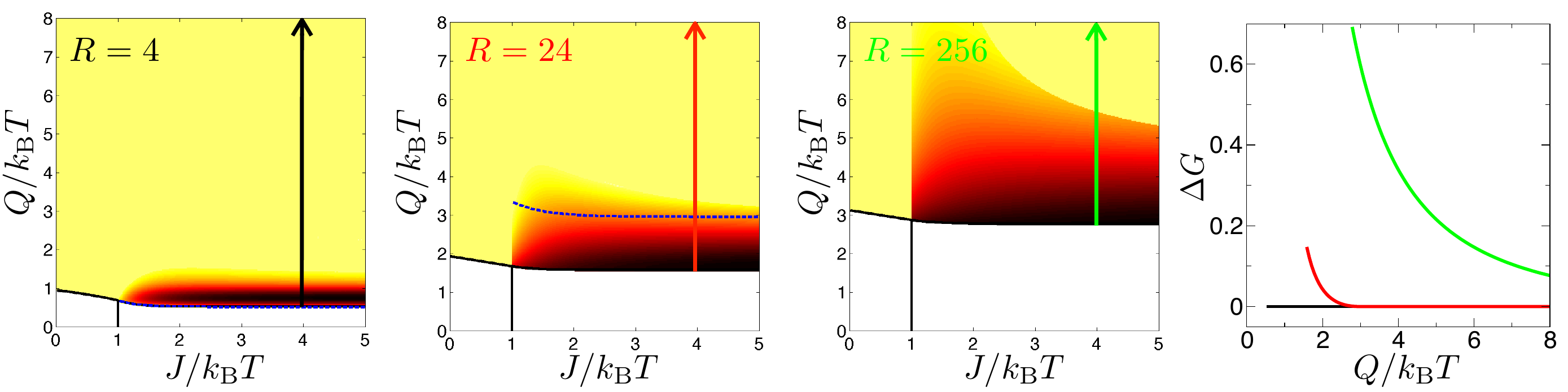} 
\caption{\label{fig6} Summarizing the shaping of crystallization landscapes by microscopic parameters. We show mean-field phase diagrams for three values of $R$, shaded according to where direct (light) and indirect pathways (dark) are favored. Right panel: barrier height $\Delta G$ between liquid and crystal phases, as a function of $Q$, for $J=4\, \kt$ (vertical lines on the three diagrams). The change of crystallization driving force from direct to indirect can both preempt (large $R$) and follow (small $R$) the loss of liquid metastability.}
\end{figure*}

The analytic theory allows a comprehensive survey of parameter space. Increasing $R$, the number of accessible orientational states, has the effect of increasing the bulk free energy barrier between crystal and liquid phases (Fig.~\ref{fig5}), a trend reflected in simulations of finite-size clusters (Fig.~\ref{fig3}). Fig.~\ref{fig6} shows pathway diagrams for three particles with different value of $R$, which display similar trends (though a stronger specific interaction is needed to effect a change of pathway for large $R$). Finally, Fig.~\ref{figa6} shows the relation between the $J,Q$ phase diagram and the conventional temperature-density one. The analytic estimate of pathway is a useful starting point for directing simulations, but it contains no dynamical information: we have seen in simulations that sluggish rotational dynamics can render the actual crystallization mechanism different to the thermodynamically-preferred one.

\section{Conclusions}

We have identified the microscopic controls of crystallization in a model of anisotropic particles that can form both disordered and crystalline phases. Our work complements previous studies that reveal how changes of intensive parameters, such as temperature~\cite{duff2009nucleation} and pressure~\cite{desgranges2007controlling}, can change crystallization pathways in model systems. Although rules of thumb are of limited use in predicting our model's crystallization pathway, a combination of simulation and analytic theory reveals that a liquid phase is likely to be seen prior to crystallization if 1) particles rotate sluggishly; 2) if particles must align precisely in order to crystallize (i.e. if $R$ is large); and 3) if the specific attraction $Q$ is just strong enough to render the crystal stable (but not so strong as to render small droplets of crystal lower in free energy than droplets of liquid of similar size). While our model is idealized, and certainly does not capture the detailed microscopics of the way real particles interact, it does show that the type of complex crystallization pathway observed in many experiments emerges as soon as one assigns to a particle a translational degree of freedom and a rotational one. Further, while we expect differences in the details of particle interactions in two- and three dimensions, the correspondence of mean field theory (which tends to work better the higher the dimension) and simulation results, and related dynamic `crossovers' seen in three-dimensional models~\cite{duff2009nucleation}, suggest that the results of this study are not limited to two-dimensional systems. With this in mind, we conjecture that the observations made here might be used as a starting point to guide experiments. For instance, they suggest that one could control the crystallization pathway of a protein if one could devise ways of altering, independently, that protein's nonspecific attraction `$J$' (e.g. using PEG) and specific attraction `$Q$' and `$R$' (e.g. through mutation or using multivalent salt~\cite{zhang2011novel}). One might also design anisotropic nanoparticles of specific rotational and translational mobilities in order to encourage a direct or indirect crystallization pathway. Finally, we used the `pathway diagram' of Fig.~\ref{fig4} to select interesting regions of phase space for our simulation work: analytic treatment of other models might furnish similar microscopic `maps' to guide simulation studies.

\section{Acknowledgements.} We thank Jim DeYoreo, Tom Haxton, Rob Jack, Daphne Klotsa and Dina Mirijanian for comments on the manuscript, and Daan Frenkel, Phill Geissler, Lutz Maibaum and Will McKerrow for discussions. L.O.H. was supported by the Center for Nanoscale Control of Geologic CO$_2$, a U.S. D.O.E. Energy Frontier Research Center, under Contract No. DE-AC02--05CH11231. This work was done at the Molecular Foundry, Lawrence Berkeley National Laboratory, supported under the same DOE Contract No. We thank NERSC for computational resources.

\appendix 

\renewcommand{\theequation}{A\arabic{equation}}
\renewcommand{\thefigure}{A\arabic{figure}}

\section{Simulation details}

We simulated our model using the following grand canonical Metropolis Monte Carlo procedure. This procedure effects a diffusive dynamics and assumes the substrate to be in contact with a thermal bath and a particle bath (i.e we assume the substrate to be in contact with bulk solution). We select at random a lattice site. If that site is occupied by a particle then with probability $p_{\rm del} \leq 1$ we attempt to delete its occupant. We accept this deletion with probability $P_{{\rm  delete}}={\rm min}\left(1,\frac{1}{p_{\rm del} R} \exp \left(-\beta \Delta E -\beta \mu\right) \right)$, where $\beta \equiv 1/(k_{\rm B} T)$ and $\Delta E$ is the change of interaction energy following the proposed deletion. With probability $1-p_{\rm del}$ we instead attempt to change the particle's orientation by $\pm 1$ unit, modulo $R$ (we assume particles to rotate in a plane, and so orientation $R$ neighbors orientation $1$). We accept changes of rotation with probability $P_{{\rm  rotate}}={\rm min}\left(1, \exp \left(-\beta \Delta E \right) \right)$. If the chosen site is vacant then we attempt to occupy it with a particle whose orientation is chosen randomly. This attempt succeeds with probability $P_{{\rm  insert}}={\rm min}\left(1,p_{\rm del} R \, \exp \left(-\beta \Delta E +\beta \mu\right) \right)$. The factors of $p_{\rm del}$ in insertion and deletion rates are required to preserve detailed balance: insertions are {\em always} attempted if a lattice site is vacant, but deletions are attempted only with probability $p_{\rm del} \leq 1$ if a lattice site is occupied. The factors of $R$ are present for a similar reason: insertion of a particle of a given orientation is attempted with probability $1/R$, but proposing the reverse of that particular insertion occurs with unit probability. We used a lattice of $N=(100)^2$ sites, periodically replicated in each direction.

The basic rate for particle translations (adsorptions and desorptions) is $\sim p_{\rm del}$. The basic rate for particle rotations has two contributions: the first scales as $(1-p_{\rm del})/R^2$, and comes from explicit rotation moves; the factor of $R^2$ accounts for the characteristic time to visit $R$ rotational states. The second contribution scales as $p_{\rm del}/R$, and comes from explicit translations (particles attach to the substrate with randomly-chosen orientations, and so removal and reattachment of a particle allows an effective sampling of its orientation). The first mode of orientation-sampling is most effective in the bulk of a liquidlike cluster, while the second mode operates most readily at the surface of a cluster (where particle detachments are most frequent). There is therefore no constant effective rate at which a cluster explores the configuration space $(N,N_{\rm c})$ shown in Fig. 1. We report values of a parameter $r \equiv (1-p_{\rm del})/p_{\rm del}$, the relative rate of proposing a rotation or translation. Our scheme does not account for in-plane particle diffusion, focusing instead on mass transport from the bulk.

We set $\mu = -Jz/2-k_{\rm B} T \ln R$, where $z=4$ is the coordination number of the lattice. Our choice of $\mu$ ensures that, to a mean-field approximation, liquid and vapor phases are equal in free energy. The contribution $-Jz/2$ is the usual Ising model term. The term $-k_{\rm B} T \ln R$ penalizes particles relative to vacancies by an amount that exactly compensates the entropy difference between particles and vacancies. This choice is motivated by the fact that we consider our simulation protocol to reflect the diffusion of material to and from the substrate, with no change of that material's rotational freedom, rather than to model its creation or destruction (which {\em would} be accompanied by creation or destruction of rotational entropy).  In simulations, orientational correlations in the liquid lower its free energy below that of the vapor by an amount that decreases with increasing $R$.

In umbrella sampling simulations, cluster size $N$ and crystallinity $N_{\rm c}$ were constrained using harmonic bonds of spring constant 0.2 within windows of width 5 (particles).   Dynamical trajectories were sown, grown and harvested using the original forward flux sampling (FFS) algorithm \cite{Allen:2006} and the aimless shooting transition path sampling algorithm~\cite{Peters:2006,Peters:2007}. FFS simulations used an interval of 5 between interfaces, 1000 starting configurations in the initial basin and $10^4$ trials per interface.

\begin{figure*}[t]
\includegraphics[width=0.6\linewidth]{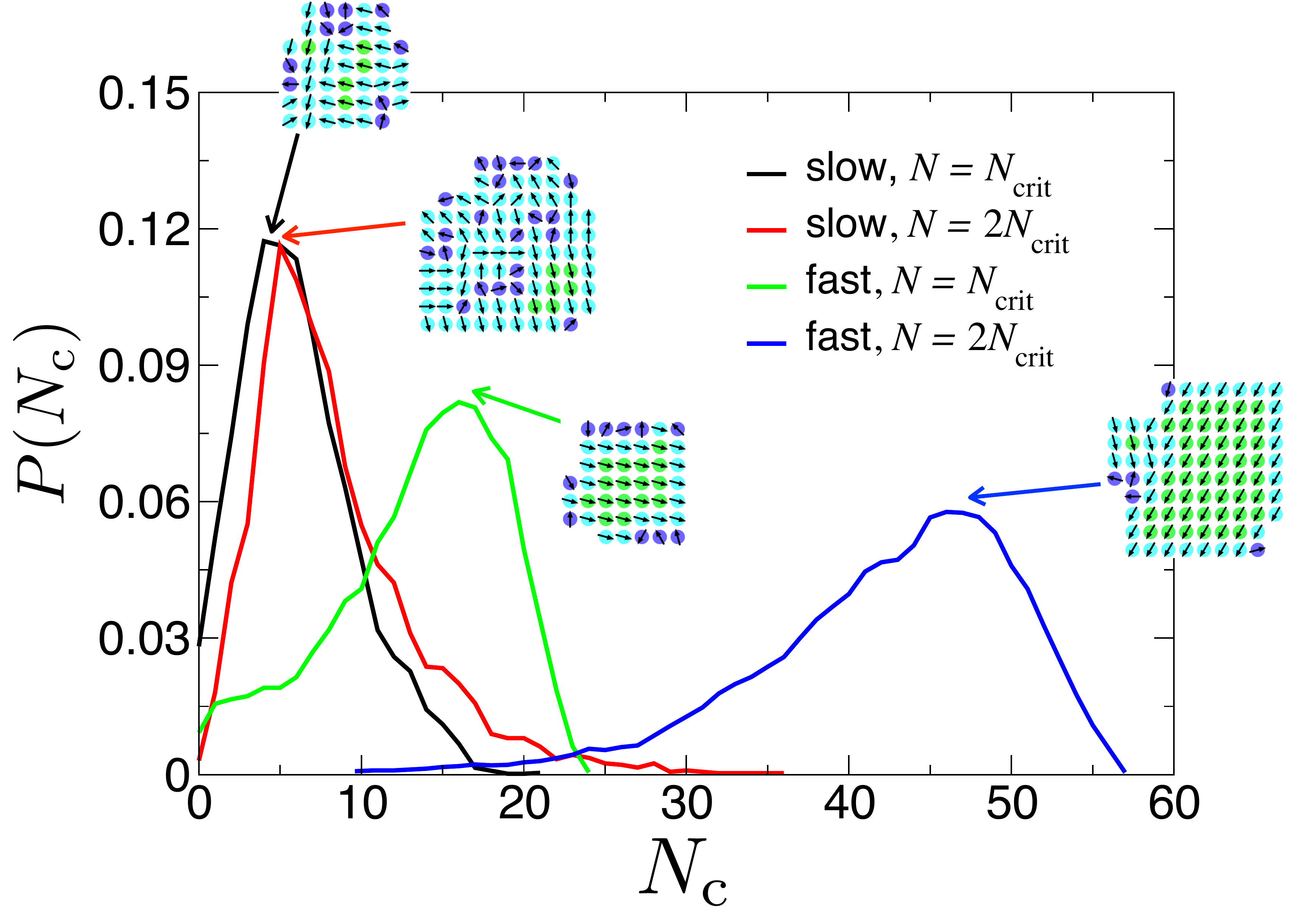} 
\caption{\label{figa3} Distributions of the number of crystalline particles in a cluster, $N_{\rm c}$, from dynamical trajectories at phase point C (see Fig. 1, main text). Distributions correspond to clusters at criticality, $N_{\mathrm{crit}}$, and twice the critical size, $2N_{\mathrm{crit}}$. For sluggish rotation rates, distributions are peaked at low values of $N_{\rm c}$: on average the growing nucleus is liquidlike. For sufficiently rapid rotation rate the direct pathway is preferred. In all cases, trajectories can be observed that buck the trends shown (note the tails of each distribution).} 
\end{figure*}

\begin{figure*}[t]
\includegraphics[width=\linewidth]{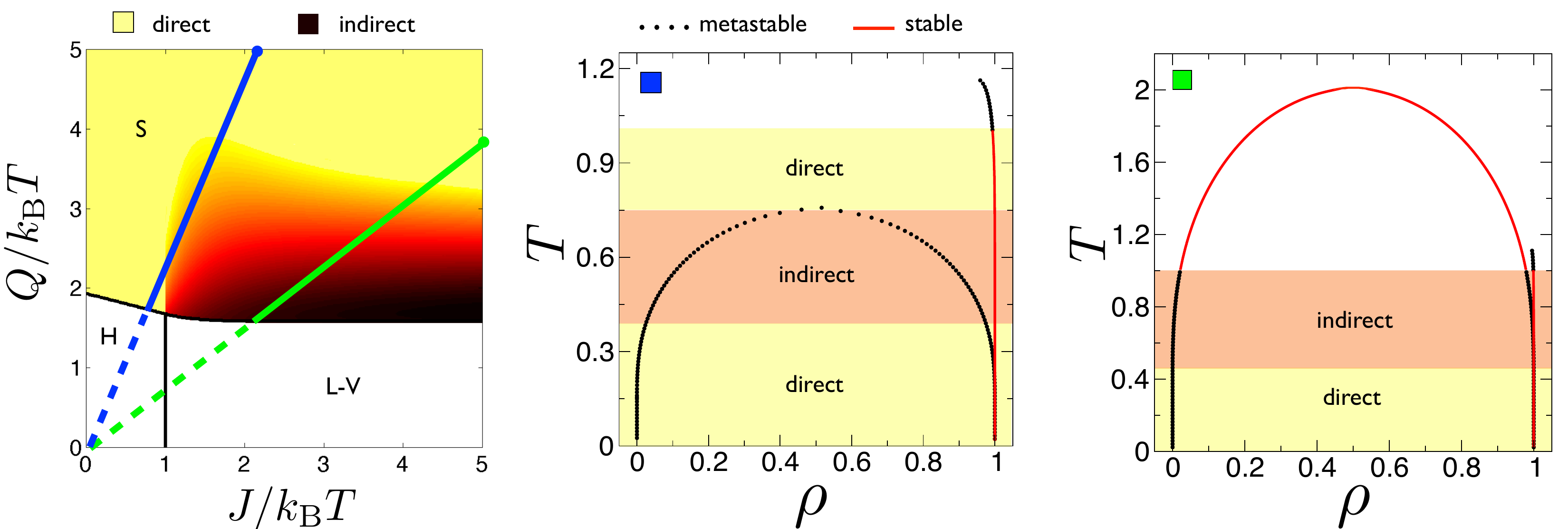} 
\caption{\label{figa6} Supplement to Fig.~\ref{fig4}: relation of the $J,Q$ phase diagram to the more conventional $T,\rho$ one. For two different varying-temperature lines ($J/Q=2.3$ and $0.8$) along our mean-field phase diagram (left, $R=24$), we show phase diagrams in the density ($\rho$)-temperature ($T$) plane. These phase diagrams resemble qualitatively those of a protein (middle) or argon (right), with their respective stable- and metastable fluid-fluid demixing critical points~\cite{PhysRevLett.77.4832}. We see that depending on where one lies in parameter space, changing temperature can change the thermodynamically preferred crystallization pathway in a complicated way.}
\end{figure*}

\section{Mean-field theory}

The energy function of our model is $\cal{H}=\sum_{i=1}^N \left(\frac{1}{2} \sum_j U_{ij}-\mu n_i\right)$, where $j$ runs over the $z=4$ nearest neighbors of $i$, and $\mu$ is a chemical potential. The variable $n_i$ is 0 if lattice site $i$ is vacant, and is 1 if it is occupied. The pairwise interaction $U_{ij}$ is
\beq
U_{ij}= - n_i n_j \left(J+Q \delta(s_i,s_j)-Q\tilde{\delta}(s_i,s_j)\right), \nonumber
\eeq
where $s_i=1,2,...,R$ is the orientation of the particle at lattice site $i$. The function $\delta(s_i,s_j)$ is 1 if $s_i=s_j$ (aligned particles receive an extra energetic reward), and is zero otherwise. The function $\tilde{\delta}(s_i,s_j)$ is 1 if $s_i$ and $s_j$ are $R/2$ units different, modulo $R$ (antialigned particles receive an energetic penalty), and is zero otherwise. 

We can derive the free energy of this model in a mean-field approximation~\cite{geng2009theory,whitelam2010non} by assuming that each site feels only the thermal average of the fluctuating variables at neighboring sites. The effective interaction at a given site is to this approximation 
\bea
U_{\rm eff}= &-& J z n \av{n} \nonumber \\
  &-&Q z \sum_{q=1}^R n\,\delta(s,q) \left(  \av{n\delta(s,q)}-\av{n\tilde{\delta}(s,q)}\right), \nonumber
\eea
where we have dropped site labels. By symmetry, all but two terms in the sum over $q$ vanish. The effective free energy per site is then $f_{\rm eff} = E-TS$, where $E=\frac{1}{2} \av{U_{\rm eff}}-\mu \av{n}$ and $-TS= \kt \av{\ln P_{\rm eq}}$, where $P_{\rm eq}= {\rm e}^{-\beta \cal{H}_{\rm eff}}/{\rm Tr}\, {\rm e}^{-\beta \cal{H}_{\rm eff}}$ and ${\cal H}_{\rm eff} \equiv U_{\rm eff} - \mu n$. Thermal averages are defined self-consistently through the relation $\av{A} \equiv {\rm Tr} \left(A\, P_{\rm eq} \right)$. The trace ${\rm Tr} (\cdot) \equiv \sum_{n=0,1} \left\{ \delta_{n,1} \sum_{s=1}^R + \delta_{n,0} \right\} (\cdot )$ can be carried out by assuming, without loss of generality, the ordering direction to be $s=1$. The result is \eqq{eq1}, in which $\rho \equiv \av{n}$ is the density and $\tau \equiv \av{n \delta (s,1)}-\av{n \tilde{\delta}(s,1)}$ is the crystallinity order parameter. The latter distinguishes disordered fluid phases (for which $\tau=0$) from the ordered solid phase (for which $\tau \neq 0$).

\section{Supplementary Figures} 

\end{document}